\documentclass[graybox]{svmult}

\usepackage{natbib} 

\usepackage{mathptmx}       
\usepackage{helvet}         
\usepackage{courier}        
\usepackage{type1cm}        
\usepackage{makeidx}         
\usepackage{graphicx}        
\usepackage{multicol}        
\usepackage[bottom]{footmisc}

\makeindex             

\newcommand\araa{\textit{ARA\&A}}%
\newcommand\apj{\textit{ApJ}}%
\newcommand\apjl{\textit{ApJ}}%
\newcommand\apss{\textit{Ap\&SS}}%
\newcommand\aap{\textit{A\&A}}%
\newcommand\mnras{\textit{MNRAS}}%
\newcommand\pasp{\textit{PASP}}%
\newcommand\aj{\textit{AJ}}%

\begin{document}

\title*{The formation of low-mass stars and brown dwarfs}
\titlerunning{The formation of low-mass stars} 
\author{Dimitris Stamatellos}
\authorrunning{Stamatellos, D.} 

\institute{
Dimitris Stamatellos \at
School of Physics \& Astronomy, Cardiff University, UK, 
\email{D.Stamatellos@astro.cf.ac.uk}
}

%
%
\maketitle

\vspace{-3cm}
\abstract{It is estimated that $\sim60$\% of all stars (including brown dwarfs) have masses below 0.2~M$_{\odot}$. Currently, there is no consensus on how these objects  form. I will briefly review the four main theories for the formation of low-mass objects: turbulent fragmentation, ejection of protostellar embryos, disc fragmentation, and photo-erosion of prestellar cores. I will focus on the disc fragmentation theory and discuss how  it addresses critical observational constraints, i.e. the low-mass initial mass function, the brown dwarf desert, and the binary statistics of low-mass stars and brown dwarfs. I will examine whether observations  may be used to distinguish between different formation mechanisms, and give a few examples of systems that strongly favour a specific formation scenario. Finally, I will argue that it is likely that all mechanisms may play a role in low-mass star and brown dwarf formation.}

\section{Introduction}

Star formation is  a process that produces objects with a wide range of masses: from a few Jupiter masses up to a few hundred solar masses. The initial mass function (IMF), i.e. the distribution of stellar masses at birth, is relatively well constrain down to  $\sim20-30$~M$_{\rm J}$, but it is uncertain at lower masses due to the difficulty in observing low-mass objects. The IMF can be approximated either by power laws \citep{Kroupa:2001a} or by a log-normal distribution \citep{Chabrier:2003a, Chabrier:2005a}. Most stars in the Galaxy have low-mass; more than $\sim 60\%$ of all stars (including brown dwarfs) have masses below 0.2~M$_\odot$ \citep[e.g. using the  Kroupa IMF;][]{Kroupa:2001a}. 

The low-mass end of the IMF is populated by three types of objects: low-mass hydrogen-burning stars, brown dwarfs, and planets. The distinction between these types of objects is done solely on their masses: stars can sustain H-burning ($m>80$~M$_{\rm J}$), brown dwarfs cannot sustain H-burning but they can burn  deuterium (13~M$_{\rm J}<m<80$~M$_{\rm J}$), and planets ($m<13$~M$_{\rm J}$) cannot sustain deuterium burning. However, it is possible that all these type of objects may form similarly, i.e. by gravitational fragmentation of gas. Indeed there is no theoretical reason for fragmentation to stop functioning at the H-burning or the D-burning limit. The theoretical minimum mass for gas fragmentation is the opacity limit at $\sim~1-5$~M$_{\rm J}$  \citep[e.g.][]{Whitworth:2006a}.

Brown dwarf and low-mass star formation requires high densities. 
The critical mass that a lump of gas needs to have in order to collapse is 
\begin{equation}
M_{\rm JEANS}=\frac{4\pi^{5/2}}{24}\frac{c_s^3}{\left(G^3\rho\right)^{1/2}}\, 
\end{equation}
where $c_s$ is the sound speed, and $\rho$ is the density of the lump.
Thus, assuming that this lump will form a brown dwarf then $M_{\rm JEANS}<80~{\rm M_J}$, from which we obtain that $\rho>10^{-16}{\rm g\ cm^{-3}}$, and $R\sim 500$~AU.  Brown dwarf formation theories attempt to either explain how these high densities are attained (e.g. in converging turbulent flows or discs) or they circumvent this requirement by forming brown dwarfs as failed stars (e.g. by ejection or photo-evaporation).

\section{Turbulent fragmentation of molecular clouds}
In the turbulent fragmentation model the high densities that are required for the formation of low-mass stars and brown dwarfs are achieved  in converging turbulent flows \cite{Padoan:2004a,Hennebelle:2008c, Hennebelle:2009b}.  

The theory reproduces the IMF and predicts that it depends on various parameters, such as the global Mach number and the thermodynamics of gas. However, the dependance is rather small for Galactic environments  and only at very low-masses, for which current observations are incomplete \citep[e.g. see review by][]{Bastian:2010a}. Nevertheless,  the theory can in principle be tested by observations in extreme environments and with more sensitive observations of the low-mass end of the IMF. 

Turbulent fragmentation has  difficulty in explaining the formation of low-mass binaries. Random pairing of stars with masses drawn randomly from the IMF does not seem to reproduce the properties of low-mass binaries \cite{Reggiani:2011a}.  Furthermore, the theory predicts the existence of gravitationally bound brown dwarf-mass cores, which have not been observed in the large numbers that are expected. However, there are examples of such cores, e.g. the  pre-brown dwarf core Oph~B-11~\cite{Andre:2012a}. 

\section{Ejection of protostellar embryos}

In this theory the collapse of a prestellar core results in the formation of a few ($\stackrel{>}{_\sim}3$) objects. Inevitably, as these objects dynamically interact with each other, the lowest mass object(s) will be ejected from the system. The mass growth of the ejected objects stops as soon as they  leave their parent core; therefore if these ejection(s) happens early on, then the mass of an ejected object would be low, even in the brown dwarf-mass regime \cite{Reipurth:2001a, Bate:2002a,Goodwin:2004a}.  

Initially it was thought that these ejections mean that the velocity dispersion of brown dwarfs seen in clusters should be higher than the velocity dispersion of stars; however, later it was shown that both populations have similar velocity dispersions \citep[e.g.][]{Bate:2003b} as low-mass stars are also frequently ejected in these type of dynamical interactions. It was also argued that ejections should be rather disruptive for discs around brown dwarfs, but simulations \citep[e.g.][]{Bate:2012a} show that brown dwarf discs may survive ejections.

\section{Disc fragmentation}
Discs  form during cloud collapse due to the initial rotation and/or turbulence of prestellar cores and conservation of angular momentum. They can grow in mass as they are being fed with material from the infalling prestellar core and can become gravitationally unstable and fragment to form low-mass stars and brown dwarfs \citep[e.g.][]{Stamatellos:2009a, Attwood:2009a}. 

Numerical simulations have shown that most of the objects form by disc fragmentation are brown dwarfs, but low-mass stars are also likely to form \cite{Stamatellos:2007c, Stamatellos:2009a}. Planetary-mass objects may also form \cite{Boley:2009a} but they tend to be ejected from the system becoming free-floating planets, and thus contributing to a possibly large population of such objects \cite{Sumi:2011a}. The  IMF of the objects formed by disc fragmentation is consistent with the low-mass end of the stellar IMF. The brown dwarfs that form by this mechanism have discs with masses up to a few tens of M$_{\rm J}$ and sizes up to a few tens of AU. The model predicts that brown dwarfs that stay as  companions to Sun-like stars are more likely to have discs than brown dwarfs in the field, as discs are likely to be disrupted during ejections \cite{Stamatellos:2009a}.

The disc fragmentation model uniquely among other formation mechanisms can explain the brown dwarf  desert. This terms refers to the lack of brown dwarfs as close companions to Sun-like stars \cite{Marcy:2000a, Grether:2006a, Sahlmann:2011a}; on the contrary low-mass hydrogen-burning stars and planets are frequently observed as close companions to Sun-like stars. In the disc fragmentation model  all objects that form in the disc start off with a mass of a few M$_{\rm J}$ and they grow in mass as they accrete material from the disc \cite{Stamatellos:2009d}. The objects that form first and migrate inwards gain enough mass to become stars, whereas the ones that stay in the outer disc region increase in mass but not as much, becoming brown dwarfs. If one of the brown dwarfs from the outer disc region drifts inwards, then it is quickly ejected again into the outer disc region due to  dynamical interactions with the higher-mass objects of the inner region. Therefore, the region close to the central star is populated by low-mass hydrogen-burning stars, and it is almost devoid of brown dwarfs \citep{Stamatellos:2009a}. Moreover, the inner disc region is populated by planets that form by core accretion at a later stage (after  $\sim1$~Myr). Most of the brown dwarfs are either ejected from the system becoming field brown dwarfs, or stay bound to the central star at relatively wide orbits ($\sim200-10^4$~AU); such wide-orbit brown dwarfs companions to Sun-like stars have been observed \cite{Faherty:2009a, Dhital:2010a, Zhang:2010a}.

The predictions of the disc fragmentation model regarding the properties of low-mass binaries are broadly consistent with observations. Close and wide brown dwarf-brown dwarf and brown dwarf-low-mass star binaries are common. Binaries  form  either by capture when two objects  are still in the disc of the host star, or by pairing up of individual objects are they are ejected from the disc \citep{Stamatellos:2009a}. The low-mass binary fraction predicted by the model is  $\sim 0.16$, similar to the binary fraction in young star forming regions \citep[e.g. in Chamaeleon 0.15-0.20;][]{Ahmic:2007a}. Most of the binaries have components with similar masses ($q>0.7$) in accordance with observations \citep[][note though that this may be due to observational biases, see \citep*{Janson:2012a}]{Burgasser:2007a}. Another interesting observational fact that the model reproduces is that brown dwarfs to Sun-like stars are more likely to be in binaries than brown dwarfs in the field \cite{Burgasser:2005a,Faherty:2010a}.

Can the conditions for disc fragmentation (i.e. disc size, disc mass) be realised in nature? Discs that are large enough so that their outer regions can cool fast enough (i.e. discs with radii $>70$ AU) and have enough mass to be gravitationally unstable at such radii can indeed fragment. \cite{Stamatellos:2011d} shows that even a 0.25~M$_{\odot}$-mass disc with radius of 100~AU around a 0.7 ~M$_{\odot}$-star fragments. Dynamical interactions in a cluster may also trigger fragmentation of discs with even lower masses \cite{Thies:2010a}. Observations of a small sample of young protostars did not reveal any massive early stage discs \cite{Maury:2010a}. However, \cite{Stamatellos:2011d} argue that finding early stage fragmenting discs is unlikely due to the short duration of the process (a few $10^3$~yr). Therefore a large number of young protostars needs to be observed. 

An issue that has been explored recently is whether radiative feedback from the central protostar heats and stabilises the disc suppressing brown dwarf and low-mass star formation \cite{Offner:2009a, Bate:2009a, Bate:2012a}. Most of the radiation than young protostars emit is due to accretion of material onto their surfaces.  \cite{Offner:2009a, Bate:2009a, Bate:2012a} have assumed that the accretion of material onto protostars is continuous. However, there is growing evidence that accretion of material may be episodic. FU Ori-type stars are objects whose luminosity increases for a few orders of magnitude for a few hundred years. During these events the accretion rates may be up to $10^{-4}~{\rm M}_{\odot}\ {\rm yr}^{-1}$. Additional evidence for episodic accretion comes from the luminosity problem: if one assumes a continuous accretion rate then the expected protostar luminosities are much larger than the observed ones \citep[e.g. ][]{Dunham:2010a}.
\cite{Stamatellos:2011c} and \cite{Stamatellos:2012a} have included the effects of episodic accretion in hydrodynamic simulations of star formation and have found that episodic accretion limits the effect of radiative feedback from the central protostar and allows disc fragmentation. In their model the luminosities of young protostars are high only during the episodic accretion events (for a few hundred years), but relatively low in-between episodic outbursts  (for a few thousand years); there is ample time between successive accretion outbursts during which the disc is relatively cool and therefore gravitational instabilities can grow and the disc can fragment.

The presence of magnetic fields is expected to  act against the formation of centrifugally supported discs because angular momentum is removed by magnetic effects (e.g. magnetic braking, outflows). However, it is uncertain whether  magnetic fields can totally suppress the formation of self-gravitating discs. \cite{Hennebelle:2008b} find that in the ideal MHD approximation the formation of a disc is suppressed if the magnetic field is strong enough and parallel to the rotation axis of the collapsing star-forming core. This is supported by ideal MHD simulations that include the effects of radiative transfer \citep{Commercon:2010a}.  The situation changes in resistive MHD calculations.  \cite{Machida:2011a} \& \cite{Vorobyov:2011b} find that disc formation is possible \citep[see also][]{Krasnopolsky:2010a,Li:2011a,Dapp:2012a}. More recently \cite{Seifried:2012a} and \cite{Joos:2012a} find that turbulence can offset the effect of magnetic breaking and allow the formation of discs with sizes up to 100~AU.

\section{Photo-erosion of prestellar cores} 
In this model a prestellar core with mass of a few M$_{\odot}$ is overrun by an H{\sc ii} region and it is photo-eroded \cite{Hester:1996a,Whitworth:2004a}. Therefore, only a fraction of the initial mass of the pre-stellar core forms a low-mass star or a brown dwarf.  The typical mass of an object produced by this mechanism is
\begin{equation}
\sim 0.01 {\rm M}_{\odot} \left( \frac{c_s}{0.3\ {\rm km\ s}^{-1}}\right) ^6
\left( \frac{\dot{\mathcal{N}}_{\rm LyC}}{10^{50}\ {\rm s}^{-1}} \right)^{-1/3}
\left( \frac{n_0}{10^3\ {\rm cm}^{-3}}\right)^{-1/3}\, ,
\end{equation}
where $c_s$ is the sound speed of the neutral gas of the core, $\dot{\mathcal{N}}_{\rm LyC}$ is the rate of ionising photons emitted by nearby stars, and $n_0$ is the density of the H{\sc ii} region.
This mechanism  produces brown dwarfs and  low-mass stars for a wide range of initial conditions, but it is inefficient, i.e. a rather massive pre-stellar core is needed for forming a brown dwarf. It can work only in the vicinity of OB stars (e.g. in Trapezium-like clusters); therefore, it cannot be the dominant mechanism for the formation of low -mass stars and brown dwarfs.

\section{Observational tests to distinguish between different formation mechanisms?}

{\bf IMF.} The turbulent fragmentation model reproduces the core mass function  and  the IMF, assuming a star formation efficiency \cite{Hennebelle:2009b}.  Disc fragmentation also  reproduces the low-mass end of the IMF \cite{Stamatellos:2009a}.  The simulations of \cite{Bate:2012a} that combine different formation mechanisms (turbulent fragmentation, ejection, disc fragmentation) also reproduce the IMF. Therefore, it appears that the IMF cannot be used  to distinguish between different formation scenarios. However, \cite{Thies:2007a} argue that when unresolved binaries are taken into account,  the IMF is discontinuous around the H-burning limit,  which suggests that  brown dwarfs may form differently than Sun-like stars.

{\bf Discs.} All formation models produce brown dwarfs and low-mass stars with discs.  Brown dwarfs that form in collapsing pre-brown dwarf cores will almost always form with discs  \cite{Machida:2009a}. Brown dwarfs that form in fragmenting discs of Sun-like stars are also likely to form with their own discs, but these discs may be partially or totally disrupted during the liberation/ejection process, resulting in a lower fraction of brown dwarfs with discs \cite{Stamatellos:2009a}. In the ejection scenario it is even more likely for discs to be disrupted during ejection but many still survive \cite{Bate:2012a}. Therefore, although different formation mechanisms may result in different disc fractions around brown dwarfs, observations of discs (and associated phenomena, i.e. accretion, outflows) around brown dwarfs do not favour any given formation theory.

{\bf The brown dwarf desert.} The disc fragmentation theory reproduces the lack of brown dwarf companions to Sun-like stars (in contrast to low-mass star companions and planetary companions; see Section 4, \cite{Stamatellos:2009a}). It has also been argued that angular momentum conservation of prestellar cores favours the formation of wide companions \citep{Jumper:2012a} in the turbulent fragmentation scenario.

{\bf Low-mass binaries.} The distribution of the projected separations of low-mass binaries   peaks at $\sim 3$~AU; close (sub-AU) and wide binaries ($>20$~AU) are also common (see http:www.vlmbinaries.org). Low-mass binaries tend to have components with similar masses, but this may be due to observational biases \cite{Janson:2012a}. These trends are broadly reproduced by the disc fragmentation model \cite{Stamatellos:2009a}. The simulations of  \cite{Bate:2009b,Bate:2012a} that combine different formation mechanisms (turbulent fragmentation, ejection, disc fragmentation) also reproduce the binaries properties. This suggests that dynamical interactions may play a dominant role in forming binaries and shaping their properties, and that the effect of the formation mechanism is secondary. However, it seems unlikely that brown dwarf- brown dwarf binaries that are companions to Sun-like star form by dynamical interactions in a cluster \citep{Kaplan:2012a}.

\section{Examples of different mechanisms at play}

The recently observed isolated core Oph B-11 provides an example of turbulent fragmentation working in the brown dwarf-mass regime. This is an isolated core, with   mass is $\sim15-20$~M$_{\rm J}$ and size $<460$~AU,  that is gravitationally bound \cite{Andre:2012a}. This core will probably collapse to form a single brown dwarf. Another example of brown dwarf formation by turbulent fragmentation is the young wide  brown dwarf  binary FU Tau A,B \cite{Luhman:2009c}. The components of the binary have masses $\sim5$ and $\sim15$~ M$_{\rm J}$. This pair is located in the Barnard 215 dark cloud and there is no higher-mass star nearby, in the disc of which the pair could have formed and then ejected.

The HL Tau system provides a possible example of disc fragmentation. The system consists of a star with mass $\sim 0.3$~M$_{\odot}$ that has a disc with mass  $\sim 0.2$~M$_{\odot}$ and radius of $>100$~AU.  1.3 cm VLA observations  have revealed the presence of a condensation with mass $\sim 14$~M$_{\rm J}$ at distance of  $\sim 65$~AU from the central star  \cite{Greaves:2008a}. Simulations have shown that disc fragmentation may be responsible for forming this object  \cite{Greaves:2008a}. The planetary system of HR8799 \cite{Marois:2008a,Marois:2010a} also provides a possible example of disc fragmentation. This is  4-planet system with four giant planets (each one with mass $\sim 10~{\rm M_{\rm J}}$) on wide orbits  (15-70~AU) around a 1.5-M$_{\odot}$ A-type star. These planets are unlikely to have formed by core accretion as they are relatively massive and orbit at large distances from the central star.

\section{Conclusions}
All formation mechanisms are probably feasible in nature and are likely to produce brown dwarfs and low-mass stars, even working in conjunction with each other. This is evident in the hydrodynamic simulations of cluster formation \citep[e.g.][]{Bate:2009b, Bate:2012a}. In simulations with no radiative transfer  $\sim75$\% of brown dwarfs form by disc fragmentation, and 25\% in dense filaments caused by turbulence \citep[with ejections happening in both cases;][]{Bate:2009b}. In simulations with radiative transfer  $\sim$20\% of brown dwarfs form by disc fragmentation, and $\sim80$\% in dense filaments \citep{Bate:2012a}. However, these simulations do not include the effects of episodic accretion that promote disc fragmentation \cite{Stamatellos:2012a}. Thus,  the actual fraction of brown dwarfs formed in discs could be between the previously mentioned limits (i.e. $20-75$\%). Therefore, it is important for star formation theories to determine the fraction of low-mass stars and brown dwarfs that form with different mechanisms and in what extent these fractions are  affected by the environment and the physical processes involved in star formation. 


\end{document}